\documentclass[pmlr,twocolumn,10pt,letterpaper]{jmlr} 

\usepackage{booktabs}

\usepackage{siunitx}
\usepackage[switch]{lineno}
\usepackage{xcolor}
\usepackage{hyperref}

\usepackage{amsmath,amssymb,amsfonts}
\usepackage{algorithmic}
\usepackage{graphicx}
\usepackage{textcomp}
\usepackage{xcolor}
\usepackage{soul}
\usepackage{booktabs}
\usepackage{tabu}

\theorembodyfont{\upshape}
\theoremheaderfont{\scshape}
\theorempostheader{:}
\theoremsep{\newline}

\jmlrvolume{LEAVE UNSET}
\jmlryear{2024}
\jmlrsubmitted{LEAVE UNSET}
\jmlrpublished{LEAVE UNSET}
\jmlrworkshop{Conference on Health, Inference, and Learning (CHIL) 2024} 

\title[Predicting Hand Motion Intentions with Multimodal Data]{A Machine Learning
Approach for Predicting Upper Limb Motion Intentions with Multimodal Data in Virtual Reality
\\}

\author{
\Name{Pavan Uttej Ravva} \Email{ravva@udel.edu}\\
\Name{Pinar Kullu} \Email{pkullu@udel.edu}\\
\Name{Mohammad Fahim Abrar} \Email{fahim@udel.edu}\\
\Name{Roghayeh Leila Barmaki} \Email{rlb@udel.edu}\\
\addr University of Delaware, Newark, DE, USA
}

\begin{document}

\maketitle

\begin{abstract}
Over the last decade, there has been significant progress in the field of interactive virtual rehabilitation. Physical therapy (PT) stands as a highly effective approach for enhancing physical impairments. However, patient motivation and progress tracking in rehabilitation outcomes remain a challenge. This work addresses the gap through a machine learning-based approach to objectively measure outcomes of the upper limb virtual therapy system in a user study with non-clinical participants. In this study, we use virtual reality to perform several tracing tasks while collecting motion and movement data using a KinArm robot and a custom-made wearable sleeve sensor. We introduce a two-step machine learning architecture to predict the motion intention of participants. The first step predicts \textbf{reaching task segments} to which the participant-marked points belonged using gaze, while the second step employs a Long Short-Term Memory (LSTM) model to predict \textbf{directional movements} based on resistance change values from the wearable sensor and the KinArm. We specifically propose to transpose our raw resistance data to the time-domain which significantly improves the accuracy of the models by 34.6\%.
To evaluate the effectiveness of our model, we compared different classification techniques with various data configurations. The results show that our proposed computational method is exceptional at predicting participant's actions with accuracy values of 96.72\% for diamond reaching task, and 97.44\% for circle reaching task, which demonstrates the great promise of using multimodal data, including eye-tracking and resistance change, to objectively measure the performance and intention in virtual rehabilitation settings.
\end{abstract}
\paragraph*{Data and Code Availability}
This paper uses two types of data: gaze data collected from the HTC Vive Pro Eye tracker and resistance data collected from a wearable sleeve sensor made of carbon nanotubes. Detailed explanations of the data collection process are provided in later sections. 
The data and code generated during the current study are publicly accessible via the following 
\mbox{\textit{\href{https://github.com/pavanravva/A-Machine-Learning-Approach-for-Predicting-Upper-Limb-Motion-Intentions-with-Multimodal-Data.git}{GitHub repository link}}}.

\paragraph*{Institutional Review Board (IRB)}
The research has been sanctioned by the Institutional Review Board (IRB) of the University of Delaware, ensuring compliance with all ethical norms concerning research with human participants. This encompasses the protection of participant confidentiality and the mitigation of any potential harm. Ethical approval was obtained on Feb 3, 2022 under protocol Number~1982585-1, and applies for the duration of the project.

\section{Introduction}
\label{sec:intro}
Over the past years, the use of virtual reality (VR) has increased significantly. VR involves various technologies that create an engaging, simulated digital world. Users can interact within this environment, which responds to their movements, fostering a sense of presence in the virtual environment.
Moreover, there have been substantial developments in these interactive virtual settings specifically designed for the motor skills rehabilitation. These improvements are especially beneficial for patients with cognitive issues and those undergoing orthopedic rehabilitation.

In addition, technology advancements in robotics and artificial intelligence progress the rehabilitation process for upper extremity patients \citep{araujo2020impact}. Patients impacted by from stroke need to perform reaching and stretching movements to regain mobility in the upper limb. The VR setup for the upper extremity offers benefits like an increase in patient engagement in therapy, easy setup for home, and no distraction and disturbance from the environment \citep{baron2021enjoyable, baron2023virtual}. Robotics rehabilitation also shows a huge improvement in terms of the recovery process. Despite its potential, the research on combining VR and the endpoint robot is limited \citep{mubin2019exoskeletons, tarnita2022analysis}.

In this work, we utilized a framework for upper extremity rehabilitation that integrates both VR and endpoint robotics. The study involved 16 participants, each engaging in two distinct reaching tasks within the framework: drawing shapes such as a circle and a diamond. The participants were required to hit each of the 40 points arranged in the shapes of a diamond or circle.

This setup was crucial for evaluating their ability to execute planar movements across multiple joints, offering insights into the smoothness and precision of their reaching trajectories \citep{kwakkel2019standardized}. We have extracted two different types of data using this framework: gaze data and resistance values. The gaze data was collected from the VR headset which provides information about the visual focus throughout the task. The resistance data was captured from nano-composite sensors embedded in the wearable sleeve. These values were recorded during muscle flexion and extension while performing reaching tasks with the sensor effectively converting the physical energy exerted during muscle movement into measurable electrical signals \citep{saggio2015resistive}. The above-described framework used for data collection exhibits significant advancements in the field of rehabilitation and improvement of motor function. 

In this study, we used these two types of data, gaze and resistance values, to predict the motion intentions of participants. To assess the effectiveness of our proposed approach, we applied different machine learning models with various data setups and compared the results with our proposed architecture. This comparative analysis provides insights into the performance and advantages of our proposed methodology over alternative approaches. The study on predicting the motion intentions of participants allows therapists to identify specific motor deviations. Therapists can use this information to tailor rehabilitation exercises, potentially leading to a more efficient and effective rehabilitation process. 

The contributions of this paper include:

\begin{itemize}
    \item Introducing a machine learning framework for prediction of the segmentation of trajectory shapes and directional movement.
    \item Integrating and leveraging multiple data modalities, such as gaze data and resistance measurements from sleeve sensors.
    \item Converting resistance measurements obtained from wearable sleeve sensors into time-domain features to attain higher accuracy.
    \item Incorporating KinArm robot data for analyzing motion intention, thereby enhancing the efficacy of predictive models for potential upper extremity rehabilitation tasks.

\end{itemize}
This study is organized as follows: Following the introduction, the related works are discussed in Section~\ref{section_rw}. The methodology of the work is explained in Section~\ref{section_methods}. The results are presented in Section~\ref{section_results}. The discussion, including implications, limitations, and future research directions, is in Section~\ref{section_discussion}. Lastly, the conclusion is presented in Section~\ref{section_conclusion}

\section{Related Work}
\label{section_rw}

In this section, we review some previous work related to virtual reality and robotics for upper extremity rehabilitation and the use of machine learning for upper limb motion prediction.

\subsection{VR Therapy and Robots}
\label{section_vr_therapy_and_robotics}

Recent technological advancements have significantly increased the demand for human-machine interaction, particularly in the realm of robotics. This surge in interest has notably impacted the field of rehabilitation, leading to a deeper understanding of motor control. This work focuses on developing a better machine learning algorithm that can precisely analyze the motion intentions. Robots can analyze the movement of limbs help in controlling the motions of the joint and provide precise motion direction for effective therapy \citep{durandau2019voluntary}. When compared with normal physical therapy, assistive robots result in a more positive outcome in rehabilitation because patients can train for longer periods \citep{daunoraviciene2018effects}. Robots can also assist in improving motor control and restoring neurological functions \citep{dixit2019effectiveness}. \cite{kim2023three} developed an electromagnetic robot for the rehabilitation of the upper extremity in sub-acute stroke patients. Its real-time position tracking feature allows for active flexion and extension, irrespective of the hand’s position. The researchers found that this robotic rehabilitation system aided stroke patients in improving their hand-motor function. This system also helps in tracking the motion in real-time and get feedback from it. 

Virtual reality-based rehabilitation in stroke patients demonstrates improvements in patient motivation, functional recovery, motor function, and dual-task performance \citep{aderinto2023exploring}. \cite{anwar2021novel} conducted a study on 68 stroke patients to compare VR training and physical therapy, revealing that VR was more effective in improving balance and lower limb function. \cite{vibhuti2023efficacy} presented a systematic review of VR therapy for various neuromotor impairments such as stroke, cerebral palsy, spinal cord injury, and parkinson’s disease reviewing forty-five studies. They reported that home-based VR therapy systems significantly improved the functional mobility of the patients. VR and gaming for upper extremity rehabilitation have also been shown to be effective in improving the motor recovery of the patients \citep{karamians2020effectiveness}.

\cite{walker2023virtual} specified the potential of robots and VR in improving the treatment procedure by using assistive robots. \cite{wonsick2020systematic} proposed the increase in growth of technology, the integration of both robots and the VR interface has more potential to improve rehabilitation techniques for people suffering from motor impairments.  

Wearable sensors provide continuous tracking and valuable clinical insights on physical activities such as the practice of diminished skills, thus playing an important role in rehabilitation systems \citep{dobkin2011promise}. \cite{caviedes2020wearable} proposed a wearable sensor array design for monitoring spine posture during physical therapy exercises. The sensor array consists of stretch sensors integrated into a custom-made sports garment, forming a triangular pattern resembling spinal exercise. Their pilot study showcased the potential of the wearable sensor array in monitoring both the adherence and accuracy of therapeutic spinal exercises during unsupervised home-based programs. \cite{luo2010interactive} developed an interactive VR system for arm and hand rehabilitation that includes a wearable sensor system consisting of an arm suit with an Optical Linear Encoder, two Inertial Measurement Units for tracking arm motion, and a SmartGlove for tracking finger motion. They developed two VR games that increase patient motivation and enable objective evaluation of patient progress.

The VR environment integrated with eye tracking shows improvement to the rehabilitation process \citep{gavas2018enhancing}. This data provides a clear understating of the participant's interaction with the VR environment; the eye tracking system provides proper data that can track the vision. This data is crucial in grasping and locomotion tasks \citep{desanghere2015influence}. Hence, gaze data can play a crucial role in rehabilitation and assistance of both upper and lower limb impairments \citep{cognolato2018head}. Gaze data helps to provide support to participants while doing rehabilitation tasks \citep{novak2013enhancing}. Park et al. (\citeyear{park2019feasibility})  worked on integrating eye tracking with the VR environment for vestibular rehabilitation. The integration of eye tracking in rehabilitation parallels our approach of combining the VR environment integrated with eye tracking and endpoint robotics. Park et al. observed a significant improvement in rehabilitation outcomes with the addition of eye tracking, motivating us to utilize eye tracking data in our work.
\subsection{Machine Learning for Upper Limb Motion Prediction}
\label{section_machine_learning_for_upper_limb_motion_prediction}

The integration of machine learning techniques with wearable sensor technology can enhance the efficacy of rehabilitation procedures for patients suffering from conditions requiring prolonged recovery periods, such as stroke \citep{wei2023application}. \cite{garcia2021hand} introduced a method for predicting hand kinematics for prosthesis control using high-density Electromyography (EMG) data. The actual hand kinematics were collected using the 18 sensors of CyberGlove II by CyberGlove Systems LLC. They employed artificial neural networks, which effectively predicted joint trajectories over 13 key hand movements. \cite{trigili2019detection} introduced a novel concept of using the signals by analyzing the behaviors of participants doing the reaching task. They then converted the sEMG signals into 14 time-domain features and used these features to train the machine learning algorithm to predict reaching tasks, go-forward, and go-backward movements. There research focused on basic task prediction without considering the motion trajectory or directional components. However, in our study, we focused on developing multimodal analysis to predict the motion intention.

Little et al. worked on the application of neural network in predicting the elbow motion trajectory focuses on dealing with different key elements for the model training process. Their work emphasised more on algorithm, which is rigid in individual variances \citep{little2021elbow}. Sibo Yand et al, did similar research on the topic of motion intention prediction; they proposed a machine learning model to predict the hand positions. This work suggests that modern machine learning algorithms can effectively predict the motion intention by taking the data in a sequence type \citep{yang2023learning}.

\section{Materials and Methods}
\label{section_methods}

The research project was approved by the Institutional Review Board (IRB). In this upcoming session, we describe briefly the participants, equipment used, data collection procedures, data preprocessing methods, and the study procedure. 

\subsection{Participants}
\label{section_participants}

The dataset consists of data from 16 healthy participants from the university of Delaware, with 16 participants (eight males), aged between 22 and 37 years (\textit{M}=27.38, \textit{SD}=4.13). The participants volunteered for the study and were not provided any financial compensation. Nine participants reported familiarity with video games, varying from several times a year to daily engagement, while seven had previous virtual reality experience. All participants self-reported as right-handed.

\begin{figure}[htbp]
    \centering
    \includegraphics[width=\columnwidth]{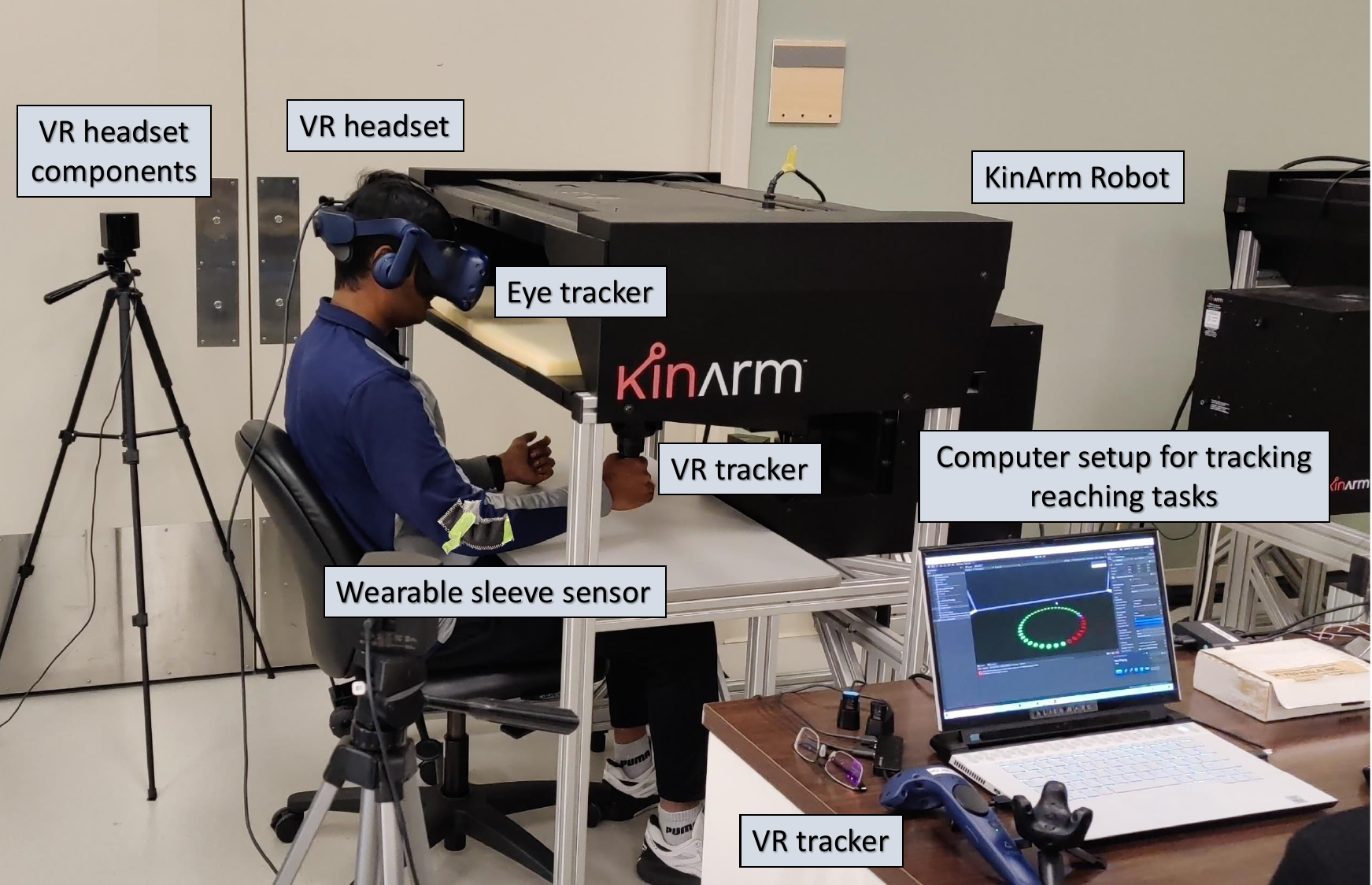}
    \caption{The upper extremity rehabilitation framework setup for data collection. The participant completes the circle task by holding the KinArm robot handle.}
    \label{fig:system-setup}
\end{figure}

\subsection{Apparatus}
\label{section_apparatus}

In this section, we explain the upper extremity rehabilitation system utilized for our experiments \citep{chheang2023immersive}. The details of the setup can be found in Figure~\ref{fig:system-setup}.
The virtual environment was developed using Unity game engine (version 2021.3.10f1).

Gaze data was collected at each hit point using the advanced eye-tracking technology of the HTC Vive Pro Eye tracker, providing detailed information on eye and head movements \citep{baron2023virtual}

The KinArm robot, featuring six degrees of freedom with a cylindrical handle for horizontal movement, was integrated into the VR setup. The KinArm model was constructed in the virtual environment with the HTC Vive tracker capturing the handle's position and inverse kinematics calculated through simulation. The KinArm robot does not provide assistive support for the participants, and in this study we didn't focus on analysis of the data from the KinArm robot. Instead, it is used for consistent hand movement of participants in 2-Dimensional space, supporting and understand the motor functions in individuals \citep{thostenson2019flexible}. KinArm also helps to provide a controlled space for motor tasks \citep{mochizuki2019movement}.

Elbow movement data was obtained using a wearable sleeve sensor made of a carbon nanotube. 
The sensor is incorporated into a knit fabric sleeve. Due to the piezo-resistive nature of the sensor, the resistance of the electrode changes due to the strain generated in the sensor during elbow flexion/extension. Participants used this sensor during reaching tasks. Data on resistance values were recorded using an Arduino-based voltage divider circuit, and data transmission occurred through Microsoft Excel. Each participant engaged in two study conditions: reaching tasks in the geometry of a circle and a diamond. 

\begin{table*}[h]
    \caption{The list of 11 time-domain features and their explanations.}
    \begin{center}
        \footnotesize
        \begin{tabular}{llp{6.5cm}}
        \toprule 
        \multicolumn{1}{c}{\textbf{Name of the Feature}} & \multicolumn{1}{c}{\textbf{Equation}} & \multicolumn{1}{c}{\textbf{Explanation}} \\
        \midrule
        Integrated Absolute Value & \(IAV = \sum_{n=1}^{N} |x_n|\) & the summation of the absolute values of the resistance values between two hit positions \\ \\
        Mean Absolute Value & \(MAV = \frac{1}{N} \sum_{n=1}^{N} |x_n|\) & the average resistance values \\ \\
        Modified Mean Absolute Value 1 & \(MMAV1 = \frac{1}{N} \sum_{n=1}^{N} w_n |x_n|\) & a modified version of MAV with a weight component (\(w_n\)) in the equation\\
        & & \(w_n = \begin{cases}
        1, & \text{if } 0.25N \leq n \leq 0.75N\\
        0.5, & \text{otherwise }
        \end{cases}\) \\ \\
        Modified Mean Absolute Value 2 & \(MMAV2 = \frac{1}{N} \sum_{n=1}^{N} |x_n|\) & a modified version of MMAV1 with a different weight function \\
        & & \(w_n = \begin{cases}
        1, & \text{if } 0.25N \leq n \leq 0.75N\\
        4n/N, & \text{if } 0.25N > n\\
        4(n - N)/N, & \text{if } 0.75N < n 
        \end{cases}\)\\ \\
        Simple Square Integral & \(SSI = \sum_{n=1}^{N} x_n^2\) & the energy of the signal \\  \\
        Variance & \(VAR = \frac{1}{N-1} \sum_{n=1}^{N} (x_n - \mu)^2\) & the variance of the resistance values, where \(\mu\) is the mean value of the resistance \\ \\
        Root Mean Square & \(RMS = \sqrt{\frac{1}{N} \sum_{n=1}^{N} x_n^2}\) & the root mean square of the resistance values \\ \\
        Waveform Length & \(WL = \sum_{n=1}^{N} |x_{n+1} - x_n|\) & the summation of the length of the waveform of the resistance signal \\ \\
        Logarithm & \(LOG = \frac{1}{N} \sum_{n=1}^{N} \log_{10}(|x_n|)\) & the average of the total summation logarithm of resistance values between consecutive hit positions \\ \\
        Skewness & \(SKEW = \frac{\frac{1}{N} \sum_{n=1}^{N} (x_n - \mu)^3}{(\frac{1}{N} \sum_{n=1}^{N} (x_n - \mu)^2)^{3/2}}\) & a measure of imbalance in the resistance data \\ \\
        Kurtosis & \(KURT = \frac{\frac{1}{N} \sum_{n=1}^{N} (x_n - \mu)^4}{(\frac{1}{N-1} \sum_{n=1}^{N} (x_n - \mu)^2)^{2}}\) & 
        a statistical equation which is used to compare the general profile of data, it reveals the outlier frequency \\
        \bottomrule
        \multicolumn{3}{p{0.8\linewidth}}{\scriptsize \(N\): the number of sensor values generated between two hitting points.}
    \end{tabular}
    \label{tab:timedomainfeatures}
    \end{center}
\end{table*}

\subsection{Data Collection}
\label{section_data-colection}

\begin{figure*}[htbp]
\floatconts
  {fig:example-plots}
  {\caption{Example plots of the measurements. The variation in resistance values captured by the sleeve sensor over time (a) for the circle task and (b) for the diamond tasks. The lowest values are observed in elbow extension and higher values for the flexion.}}
  {
    \subfigure[]{\label{fig:plot1}%
      \includegraphics[width=0.48\linewidth]{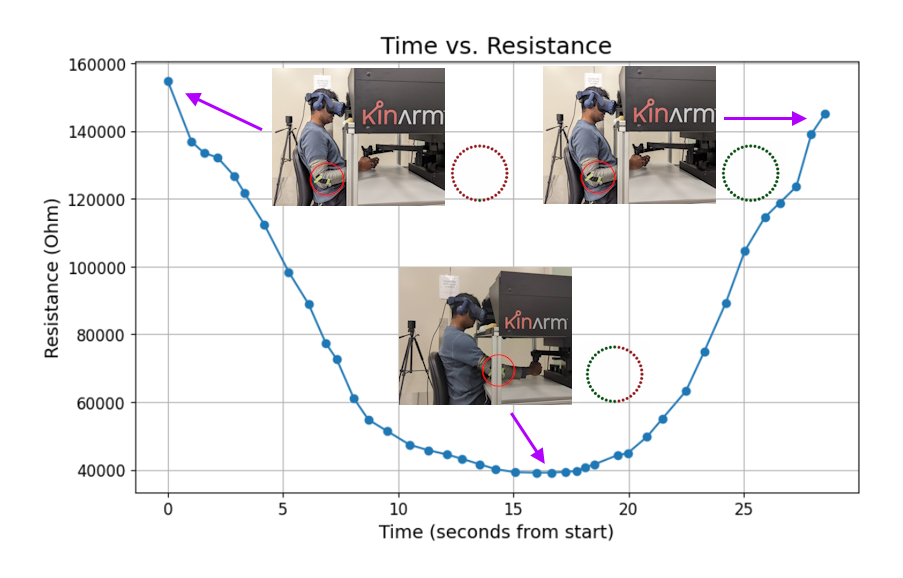}}%
    \subfigure[]{\label{fig:plot2}%
      \includegraphics[width=0.48\linewidth]{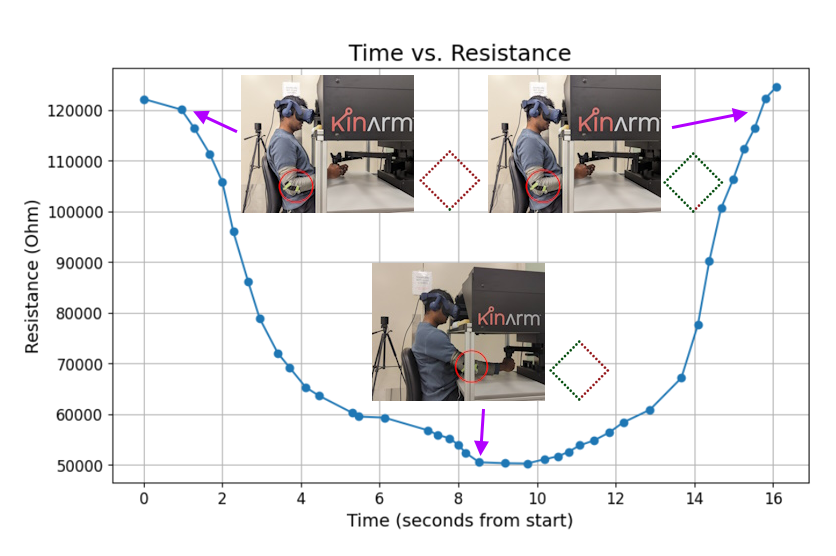}}\\[1ex]
  }
\end{figure*}

Participants were equipped with a sleeve sensor and a VR headset, and seated in front of the KinArm robot. Subsequently, participants engaged in two distinct reaching tasks: following the outline of a diamond shape and a circle. Each shape was composed of 40 points along its circumference, and participants were tasked with hitting the points sequentially. Backtracking was disallowed if a point was missed, and participants were expected to trace the shape in both clockwise and counterclockwise directions. Participants were instructed based on their group, with eight participants performing counterclockwise motions and the others executing clockwise motions.

Throughout the reaching tasks, data on resistance values were collected for each time instance, capturing the flexion/extension of muscles. Additionally, gaze data, encompassing the position and orientation of participants' heads and eyes at the moment of dot contact, was systematically recorded.

Synchronization among all the system components was ensured using packages from the Lab Streaming Layer (LSL) \citep{labstreaminglayer} for the collection of synchronized data among the sensors. LSL is an open-source system designed for the collection of time series in research experiments. Besides using LSL, manual checking of the data was necessary to ensure identical start and end times for accurate data synchronization.

Figure~\ref{fig:example-plots} provides visual examples of the data collected during the experiments. Figure~\ref{fig:plot1} and Figure~\ref{fig:plot2} illustrate the variation in resistance values recorded by the sleeve sensor throughout the reaching task for one participant, showing data for both the circle and diamond tasks, respectively. Additionally, technical setups displaying the angle of flexion and conceptual views of the task progress are included for better illustration.

\subsection{Data Preprocessing}
\label{section_data-prep}

We utilized two distinct types of data collected from the framework: gaze data for the first model and resistance data for the second model. In this section, we detail the preprocessing steps applied to the resistance data collected through carbon nanotube sensors. The objective is to transform raw resistance values into 11 time-domain features, each serving a specific purpose in characterizing upper extremity movements \citep{trigili2019detection}. For instance, the absolute value reflects overall muscle activation levels, and the root mean square relates to the force of muscle contractions. Combining all these features provides a refined overview of muscle behavior, offering a more comprehensive analysis than a single sensor value. These features are crucial for obtaining a detailed pattern of muscle behaviors and understanding the subtle aspects of human movement. Encompassing a wide range of characteristics, from the mean to the rate of signal variations, these features enhance the accuracy and robustness of analysis using machine learning.
The comprehensive list of these features and their explanations are provided in Table~\ref{tab:timedomainfeatures}.

We divided the dataset into two subsets: 80\% for training and 20\% for testing.  This split was applied to ensure that the model was trained on a sufficiently large portion of the data while also having an independent set for evaluating its performance.
Furthermore, to ensure consistency and comparability across different features, we applied regular min-max normalization to both the training and testing data. This standardized data was then used as input for subsequent machine learning models.

\begin{figure*}[htbp]
    \centering
    \includegraphics[width=0.75\textwidth]{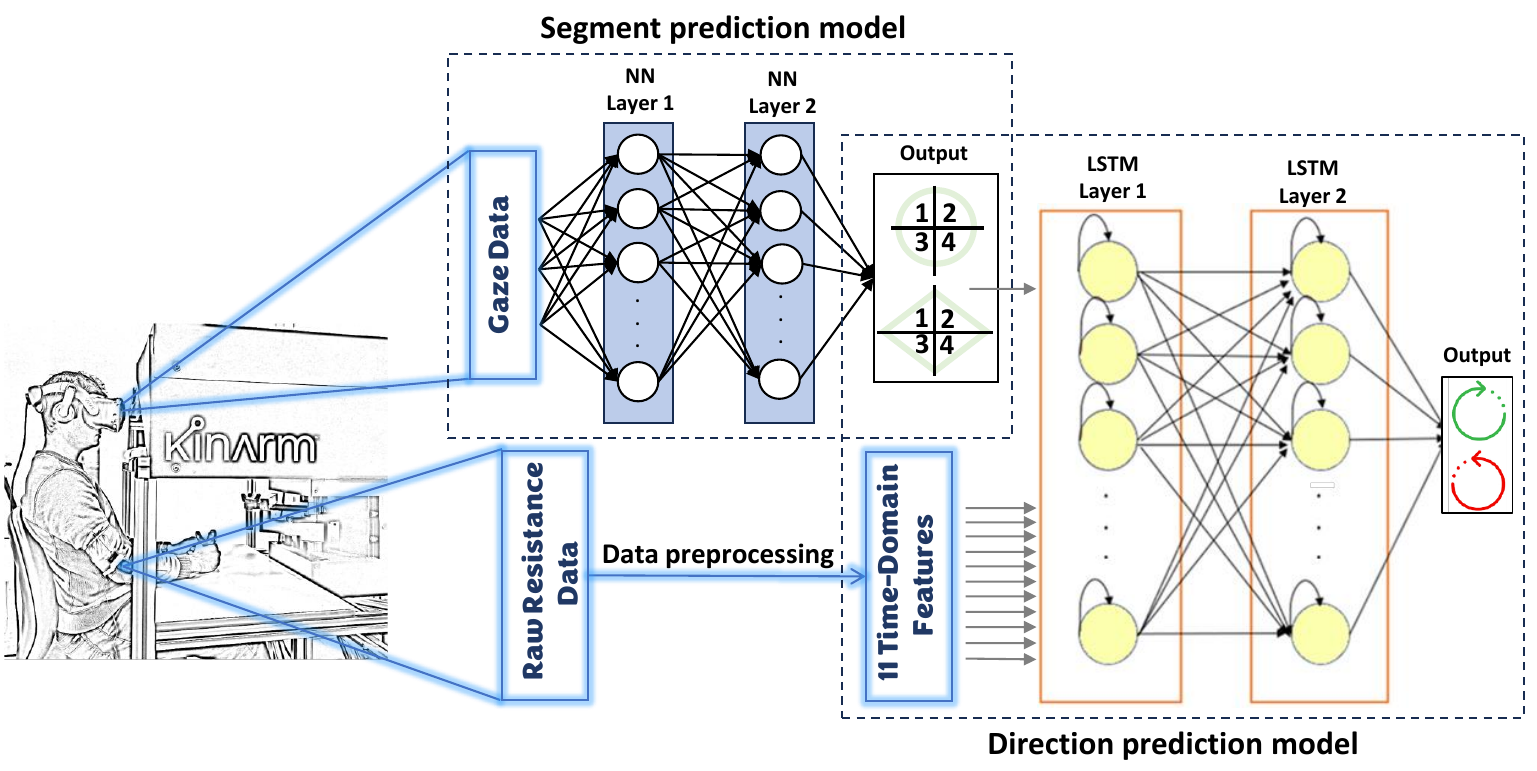}
    \caption{Overview of the two-step machine learning architecture. In the first step, we proposed a network for predicting segments using a neural network and gaze data. In the second step, we trained a model for predicting movement direction using an LSTM model with the output of the first model and 11 time-domain features.}
    \label{fig:overview_of_model}
\end{figure*}

\subsection{Study Procedure}
\label{section_study_procedure}

In this study, we propose a two-step machine learning framework designed to predict both the specific segment to which participant-marked points belong and the participant's intention to move either in a clockwise or counterclockwise direction. Figure~\ref{fig:overview_of_model} provides an overview of our two-step machine learning architecture.  In this section, we provide the details about our two models that use the data explained in Section~\ref{section_data-prep}. Our approach integrates traditional machine learning and deep learning approaches for a comprehensive understanding of motion intention.

\subsubsection{\textbf{Neural Network for Segment Prediction}}
The neural network model is designed to predict segments, which takes gaze data as input. The architecture of the neural network consists of four layers: one input layer, two hidden layers with 64 and 32 neurons, and one output layer. We used the ReLU activation function, chosen for its efficiency compared to other activation functions \citep{sharma2017activation}. The neural network was trained using the Adam optimizer with a learning rate of 0.001, 50 epochs, and a batch size of 32. We trained the input layer of the neural network by gaze data for predicting the four segments of reaching task shapes. 

\subsubsection{\textbf{Long Short-Term Memory (LSTM) Model for Direction Prediction}}
The output of our neural network model, representing the predicted segments, and 11 time-domain features of resistance data are utilized to create a specialized LSTM model. The LSTM model predicts the intended direction of movement of the participant, whether it is clockwise or counterclockwise. We chose the LSTM model for its ability to capture the temporal dependencies and patterns in the data, which are crucial in determining the direction of motion. The architecture of the LSTM model is designed as a sequential model. It consists of 40 data points arranged in sequence. The input layer takes the outputs of the neural network model and 11 time-domain features. The architecture further includes two hidden layers, each containing 50 neurons, and an output layer. We applied L2 regularization with a factor of 0.01 and used the ReLU activation function. We trained the LSTM model using the Adam optimizer with a learning rate of 0.001, 50 epochs, and a batch size of 32.  

\section{Experiments and Results}
\label{section_results}

In this section, we present the results of our experiments aimed at understanding the effect of different dataset setups on the performance of our two-step machine learning architecture. Our study focused on predicting four segments of reaching tasks in the first model and predicting the direction of movement in the second model. We explored various combinations of raw resistance data---11 time-domain features extracted from the raw resistance value, and gaze data---and investigated different machine learning models to identify the most effective approaches. We used all the data except raw resistance data as consecutive pairs of values sequentially. This sequential process involved deriving features from consecutive pairs of data points.  We present the details of all these different setups in Table~\ref{tab:benchmarks}. The size listed in the table represents the number of samples and features in each dataset, with size (samples x features). For the raw resistance data (D1), we included all 40 data points per participant, resulting in a dataset size of $640 \times 1$ ($640 = 40 \text{ data points} \times 16 \text{ participants}$). 
For the remaining, due to the sequential feature extraction process, the 40 data points per participant were reduced to 39. This resulted in a total of 624 ($39 \text{ data points} \times 16 \text{ participants}$) combined data points (samples) across all participants.

\begin{table}[htbp]
    \caption{Specifications of data setups used in the experiments.}
    \small 
    \begin{center}
    \begin{tabular}{lcc}
        \toprule
        \multicolumn{1}{c}{\textbf{Used data}} & \textbf{Name} & \textbf{Size} \\ 
		\midrule
        Raw resistance data & D1 & 640x1 \\  
        11 time-domain features & D2 & 624x11 \\  
        Gaze data & D3 & 624x24 \\  
		Output of the first model & D4 & 624x4 \\  
        The combination of D2\&D3 & D5 & 624x35 \\  
		The combination of D2\&D4 & D6 & 624x15 \\  
		The combination of D3\&D4 & D7 & 624x28 \\  
		The combination of D2\&D3\&D4 & D8 & 624x39 \\ 
	   \bottomrule		
    \end{tabular}
    \label{tab:benchmarks}
    \end{center}
\end{table}

\subsection{Segment Prediction Experiments}

\begin{table*}[h]
    \caption{The accuracy and F1 score results of the machine learning models for predicting segments in \textit{Diamond reaching task}}
    \begin{center}
    \small 
    \begin{tabu} to 0.95\linewidth {X[0.4,l] *{4}{X[c]}}
        \toprule
             & \textbf{D1} & \textbf{D2} & \textbf{D3} & \textbf{D5} \\ 
    	\midrule
        NN & 39.67 [0.325] & 49.37 [0.402] & \textbf{92.31 [0.987]} & 91.03 [0.857] \\ 
        KNN & 32.50 [0.308] & 41.00 [0.365] & 88.60 [0.897] & 85.90 [0.859] \\ 
        SVM & 38.40 [0.350] & 42.70 [0.346] & 87.20 [0.870] & 83.34 [0.830] \\ 
        LR & 31.25 [0.254] & 41.00 [0.344] & 87.12 [0.911] & 84.20 [0.935] \\
    	\bottomrule
    \end{tabu}
    \smallskip
    \begin{tabu} to 0.95\linewidth {@{} X @{}}
        \footnotesize
        All entities are in the format: Accuracy (\%) [F1 Score]. A higher Accuracy/F1 score indicates a better performance.\\
        \footnotesize
        \(D1\): Raw resistance data, \(D2\): 11 time-domain features, \(D3\): Gaze data, \(D5\): D2\&D3
    \end{tabu}
    \label{tab:model1:accuracy:diamond}
    \end{center}
\end{table*}

\begin{table*}[h]
    \caption{The accuracy and F1 score results of the machine learning models for predicting segments in \textit{Circle reaching task}}
    \small
    \begin{center}
    \begin{tabu} to 0.95\linewidth {X[0.4,l] *{4}{X[c]}}
    	\toprule    
         & \textbf{D1} & \textbf{D2} & \textbf{D3} & \textbf{D5} \\ 
    	\midrule
        NN & 38.05 [0.258] & 47.25 [0.451] & \textbf{94.87 [0.884]} & 85.06 [0.854] \\ 
        KNN & 36.25 [0.349] & 37.80 [0.577] & 87.10 [0.946] & 76.92 [0.894] \\ 
        SVM & 25.76 [0.317] & 33.57 [0.570] & 87.55 [0.797] & 84.28 [0.922] \\ 
        LR & 36.25 [0.230] & 23.08 [0.422] & 86.76 [0.823] & 84.61 [0.962] \\ 
    	\bottomrule
    \end{tabu}
    \smallskip
    \begin{tabu} to 0.95\linewidth {@{} X @{}}
        \footnotesize
        All entities are in the format: Accuracy (\%) [F1 Score]. A higher Accuracy/F1 score indicates a better performance.\\
        \footnotesize
        \(D1\): Raw resistance data, \(D2\): 11 time-domain features, \(D3\): Gaze data, \(D5\): D2\&D3
    \end{tabu}
    \label{tab:model1:accuracy:circle}
    \end{center}
\end{table*}   

\begin{table}[htbp]
    \caption{Random Guess Prediction Results - Baseline for Task Segmentation and Direction Prediction}
    \small
    \begin{center}
    \begin{tabular}{lc}
       	\toprule
        \multicolumn{1}{c}{\textbf{Prediction problem}} & \textbf{Accuracy (\%)} \\ 
    	\midrule
        Segment prediction for diamond & 26.44  \\  
        Segment prediction for circle & 25.77  \\  
        Direction prediction for diamond & 50.71  \\  
        Direction prediction for circle & 51.90  \\  
    	\bottomrule
    \end{tabular}
    \label{tab:random-guess}
    \end{center}
\end{table}

We examined various data setups (D1, D2, D3, and D5) and machine learning techniques in our first set of experiments focusing on predicting four segments of the reaching tasks. We trained models using four different machine learning methods, namely Neural Network (NN), k-Nearest Neighbors (KNN), Support Vector Machine (SVM), and Logistic Regression (LR). We used four different data setups, to explore their impact on the models' performance. 

For two reaching task shapes, diamond and circle, we listed the accuracy results and F1 scores of the trained four different machine learning models obtained with the different data setups in Table~\ref{tab:model1:accuracy:diamond} and Table~\ref{tab:model1:accuracy:circle}, respectively. All entities in these tables are presented in the format "Accuracy [F1 Score]", where the F1 Score is a metric that offers a balanced measure of a model's performance. We also compared our results with a random guess baseline. The random guess accuracy for segment prediction tasks, as shown in Table~\ref{tab:random-guess}, is around 25\% for both the Diamond and Circle. In addition to our selected NN model, the other machine learning models we experimented with also achieved better results than the baseline in most data setups. This further indicates the suitability of this problem for machine learning approaches.

After analyzing the accuracy and F1 score results presented in Tables~\ref{tab:model1:accuracy:diamond} and \ref{tab:model1:accuracy:circle}, we chose the NN model trained with gaze data. The NN model trained with gaze data demonstrated higher accuracy and F1 scores for both the diamond and circle reaching tasks, achieving 92.31\% [0.987] and 94.87\% [0.884], respectively. These results show the critical role of gaze data, as it consistently contributed to the models' best performance.  Incorporating gaze data in D3 and D5 provides supplementary cues that enhance the models' predictive abilities, thus resulting in improved performance compared to setups lacking gaze data (D1 and D2). The differences in performance between D3 and D5 highlight the possibility that integrating features from different modalities could introduce complexities or redundancies, potentially affecting the overall performance of the models.

\subsection{Direction Prediction Experiments}
In this section, we present the results of our experiments focused on predicting the direction of movements, as clockwise or counterclockwise direction. We explored all data combinations shown in Table~\ref{tab:benchmarks}, and applied multiple machine learning techniques to determine the one with the highest accuracy.
We trained models with four different machine learning methods: LSTM, KNN, SVM, and LR. 

Tables~\ref{tab:model2:accuracy:diamond} and \ref{tab:model2:accuracy:circle} present both the accuracy and F1 score results of the machine learning models for predicting the direction of movements in the diamond and circle reaching tasks, respectively. These tables provide the accuracy and F1 scores achieved by each model for different data setups. The highest accuracy and F1 score values, highlighted in bold, identify the most successful models for predicting tasks.

\begin{table*}[h]
    \caption{The accuracy and F1 score results of the machine learning models for predicting the direction in \textit{Diamond reaching task}}
    \begin{center}
    \small
    \begin{tabu}to 0.95\linewidth {X[0.2,l] *{4}{X[c]}}
        \toprule
        & \textbf{LSTM} & \textbf{KNN} & \textbf{SVM} & \textbf{LR} \\
        \midrule
        D1 & 52.50 [0.665] & 62.50  [0.576] & 50.00 [0.666] & 50.00 [0.667] \\
        D2 & 87.18 [0.796] & 94.44 [0.939] & 71.79  [0.731] & 74.36 [0.750] \\
        D3 & 76.92 [0.698] &  47.44 [0.782] & 58.97 [0.620] & 62.82 [0.701] \\
        D4 & 56.25 [0.674] & 57.50 [0.512] & 50.00 [0.666] & 50.00 [0.667] \\
        D5 & 73.08 [0.887] & 78.21 [0.798] & 79.49 [0.789] & 71.79 [0.755] \\
        D6 & \textbf{96.72 [0.946]} & 76.92 [0.833] & 75.64 [0.769] & 74.36 [0.725] \\
        D7 & 94.44 [0.939] & 43.59 [0.435] & 60.26 [0.629] & 67.95 [0.721] \\
        D8 & 96.15 [0.896] & 70.51 [0.794] & 65.38 [0.761] & 71.79 [0.769] \\
        \bottomrule
    \end{tabu}
    \smallskip
    \begin{tabu} to 0.95\linewidth {@{} X @{}}
        \footnotesize
        All entities are in the format: Accuracy (\%) [F1 Score]. A higher Accuracy/F1 score indicates a better performance.\\
        \footnotesize
        \(D1\):  Raw resistance data, \(D2\): 11 time-domain features, \(D3\):  Gaze data, \(D4\): Output of the first model, \\
        \footnotesize
        \(D5\): D2\&D3, \(D6\): D2\&D4, \(D7\): D3\&D4, \(D8\): D2\&D3\&D4\\
    \end{tabu}
    \label{tab:model2:accuracy:diamond}
    \end{center}
\end{table*}

\begin{table*}[htbp]
    \caption{The accuracy and F1 score results of the machine learning models for predicting the direction in \textit{Circle reaching task}}
    \begin{center}
    \small
    \begin{tabu}to 0.95\linewidth {X[0.2,l] *{4}{X[c]}}
        \toprule
        & \textbf{LSTM} & \textbf{KNN} & \textbf{SVM} & \textbf{LR} \\
        \midrule
        D1 & 64.7 [0.529] & 44.75 [0.473] & 52.80 [0.570] & 53.75 [0.539] \\
        D2 & 89.74 [0.720] & 51.28 [0.538] & 76.19 [0.712] & 56.41 [0.727] \\
        D3 & 61.54 [0.602] & 51.28 [0.525] & 61.53 [0.646] & 46.15 [0.502] \\
        D4 & 57.67 [0.678] & 48.75 [0.518] & 51.34 [0.472] & 51.21 [0.460] \\
        D5 & 89.74 [0.917] & 67.94 [0.679] & 42.37 [0.377] & 35.90 [0.307] \\
        D6 & \textbf{97.44 [0.923]} & 53.84 [0.577] & 74.35 [0.739] & 58.97 [0.725] \\
        D7 & 91.03 [0.894] & 61.53 [0.602] & 71.79 [0.741] & 56.41 [0.733] \\
        D8 & 96.15 [0.894] & 66.67 [0.699] & 51.28 [0.458] & 46.15 [0.337] \\
        \bottomrule
        \end{tabu}
    \smallskip
    \begin{tabu} to 0.95\linewidth {@{} X @{}}
        \footnotesize
        All entities are in the format: Accuracy (\%) [F1 Score]. A higher Accuracy/F1 score indicates a better performance.\\
        \footnotesize
        \(D1\):  Raw resistance data, \(D2\): 11 time-domain features, \(D3\):  Gaze data, \(D4\): Output of the first model, \\
        \footnotesize
        \(D5\): D2\&D3, \(D6\): D2\&D4, \(D7\): D3\&D4, \(D8\): D2\&D3\&D4\\
    \end{tabu}
    \label{tab:model2:accuracy:circle}
    \end{center}
\end{table*}

Analysis of the results shows that the LSTM model trained with the combination of D2\&D4 data is the optimal choice. This model demonstrated superior accuracy and F1 score, reaching 96.72\% [0.946] in the diamond reaching task and 97.44\% [0.923] in the Circle reaching task. The incorporation of both 11 time-domain-specific resistance features and the output of the previous segment prediction model (D2\&D4) significantly improved the accuracy and F1 scores of direction predictions for both task shapes. 

Similar to the segment prediction problem, in this context, machine learning results were compared with a random guess baseline. The random guess accuracy for direction prediction tasks, presented in Table~\ref{tab:random-guess}, is approximately 50\% for both diamond and circle tasks. Similar to the segment prediction experiments, the machine learning models for direction prediction outperformed the random guess baseline in most data setups.

\section{Discussion}
\label{section_discussion}
In this study, we propose a novel two-step machine learning architecture for predicting the motion intention of participants engaged in upper extremity rehabilitation tasks. The framework we used for extracting data integrated VR with the KinArm robot, and we utilized gaze data and resistance values to enhance the precision of motion intention predictions. The KinArm robot is primarily utilized for support in the rehabilitation, playing a role in improving the process without directly contributing to prediction analysis. 

Our first model, based on a neural network, successfully predicted the four reaching task segments. Our experiments demonstrate the importance of incorporating gaze data for accurate predictions. Eye movement information significantly improved the model's performance, highlighting the relevance of multi-modal data integration in rehabilitation studies.

Our second model, an LSTM model, accurately determined the participant's intention to move in a clockwise or counterclockwise direction. The integration of 11 time-domain features extracted from resistance data, combined with the output of the first model, achieved high accuracy in predicting bidirectional rotational behavior. The 11 time-domain features carry a wide range of statistical and temporal properties that capture the dynamic nature of resistance over time.

While some may question whether more complex models could offer better alternatives for this prediction problem, our approach highlights that, despite its simplicity, basic machine learning techniques can reveal valuable insights. By leveraging features, our models effectively capture essential information for prediction motion intentions in upper extremity rehabilitation tasks.

\paragraph{Limitations and Future Directions}
While our study provides useful insights into upper extremity movements using the proposed machine learning architecture, we must acknowledge some limitations. Firstly, the performance of our models depends on the quality and quantity of the data used, which may be impacted by sensor noise or data collection errors. Secondly, our study was focused on healthy, non-clinical participants. So, our future work will include validation with clinical populations to assess the applicability and generalizability of our approach in rehabilitation settings.

\section{Conclusion}
\label{section_conclusion}
This study presents a comprehensive approach for personalized and effective rehabilitation practices, contributing to improvements in integrating robotics, VR, and biomechanical data analysis. Our study and findings demonstrate the potential of incorporating machine learning techniques to improve the effectiveness of virtual rehabilitation assessment. For instance, our NN model achieved an impressive accuracy of 92.31\% in predicting segment tasks for the Diamond reaching task and 94.87\% for the Circle reaching task when trained with gaze data. Furthermore, our LSTM model, when trained with a combination of 11 time-domain features and the output of the segment prediction model, achieved accuracy rates of 96.72\% and 97.44\% for the Diamond and Circle reaching tasks, respectively. There are exciting possibilities for our research to explore. This research introduces an attractive application for real-time predictions, enabling robotic arms to obtain feedback from our proposed two-step model to guide participants accurately based on their motion intentions without diverting from the actual task. 

\section{Acknowledgments}
\label{section_Acknowledgments}
We wish to acknowledge the support from our sponsors, the National Science Foundation (2222663), and the National Institute of General Medical Sciences of the National Institutes of Health (P20 GM103446 and P20 GM103446E). We also express our gratitude to Drs. Vuthea Chheang, Joshua Cashback, and Erik Thostenson for their invaluable efforts in ideation, data collection, and resource sharing. We also extend our sincere appreciation to the entire research team and the study participants for their contributions. Any opinions, findings, conclusions, or recommendations expressed in this material are those of the authors and do not necessarily reflect the views of the sponsors.

\bibliography{main}

\begin{thebibliography}{35}
\providecommand{\natexlab}[1]{#1}
\providecommand{\url}[1]{\texttt{#1}}
\expandafter\ifx\csname urlstyle\endcsname\relax
  \providecommand{\doi}[1]{doi: #1}\else
  \providecommand{\doi}{doi: \begingroup \urlstyle{rm}\Url}\fi

\bibitem[Aderinto et~al.(2023)Aderinto, Olatunji, Abdulbasit, Edun, Aboderin, and Egbunu]{aderinto2023exploring}
Nicholas Aderinto, Gbolahan Olatunji, Muili~Opeyemi Abdulbasit, Mariam Edun, Gbolahan Aboderin, and Emmanuel Egbunu.
\newblock Exploring the efficacy of virtual reality-based rehabilitation in stroke: a narrative review of current evidence.
\newblock \emph{Annals of Medicine}, 55\penalty0 (2):\penalty0 2285907, 2023.

\bibitem[Anwar et~al.(2021)Anwar, Karimi, Ahmad, Mumtaz, Saqulain, and Gilani]{anwar2021novel}
Naveed Anwar, Hossein Karimi, Ashfaq Ahmad, Nazia Mumtaz, Ghulam Saqulain, and Syed~Amir Gilani.
\newblock A novel virtual reality training strategy for poststroke patients: a randomized clinical trial.
\newblock \emph{Journal of Healthcare Engineering}, 2021:\penalty0 1--6, 2021.

\bibitem[Ara{\'u}jo et~al.(2020)]{araujo2020impact}
Nuno Miguel~Faria Ara{\'u}jo et~al.
\newblock Impact of the fourth industrial revolution on the health sector: a qualitative study.
\newblock \emph{Healthcare informatics research}, 26\penalty0 (4):\penalty0 328--334, 2020.

\bibitem[Baron et~al.(2021)Baron, Wang, Segear, Cohn, Kim, and Barmaki]{baron2021enjoyable}
Lauren Baron, Qile Wang, Sydney Segear, Brian~A Cohn, Kangsoo Kim, and Roghayeh Barmaki.
\newblock Enjoyable physical therapy experience with interactive drawing games in immersive virtual reality.
\newblock In \emph{Proceedings of the 2021 ACM Symposium on Spatial User Interaction}, pages 1--8, 2021.

\bibitem[Baron et~al.(2023)Baron, Chheang, Chaudhari, Liaqat, Chandrasekaran, Wang, Cashaback, Thostenson, and Barmaki]{baron2023virtual}
Lauren Baron, Vuthea Chheang, Amit Chaudhari, Arooj Liaqat, Aishwarya Chandrasekaran, Yufan Wang, Joshua Cashaback, Erik Thostenson, and Roghayeh~Leila Barmaki.
\newblock Virtual therapy exergame for upper extremity rehabilitation using smart wearable sensors.
\newblock In \emph{Proceedings of the 8th ACM/IEEE International Conference on Connected Health: Applications, Systems and Engineering Technologies}, pages 92--101, 2023.

\bibitem[Caviedes et~al.(2020)Caviedes, Li, and Jammula]{caviedes2020wearable}
Jorge~E Caviedes, Baoxin Li, and Varun~C Jammula.
\newblock Wearable sensor array design for spine posture monitoring during exercise incorporating biofeedback.
\newblock \emph{IEEE Transactions on Biomedical Engineering}, 67\penalty0 (10):\penalty0 2828--2838, 2020.

\bibitem[Chheang et~al.(2023)Chheang, Lokesh, Chaudhari, Wang, Baron, Kiafar, Doshi, Thostenson, Cashaback, and Barmaki]{chheang2023immersive}
Vuthea Chheang, Rakshith Lokesh, Amit Chaudhari, Qile Wang, Lauren Baron, Behdokht Kiafar, Sagar Doshi, Erik Thostenson, Joshua Cashaback, and Roghayeh~Leila Barmaki.
\newblock Immersive virtual reality and robotics for upper extremity rehabilitation.
\newblock \emph{arXiv preprint arXiv:2304.11110}, 2023.

\bibitem[Cognolato et~al.(2018)Cognolato, Atzori, and M{\"u}ller]{cognolato2018head}
Matteo Cognolato, Manfredo Atzori, and Henning M{\"u}ller.
\newblock Head-mounted eye gaze tracking devices: An overview of modern devices and recent advances.
\newblock \emph{Journal of rehabilitation and assistive technologies engineering}, 5:\penalty0 2055668318773991, 2018.

\bibitem[Daunoraviciene et~al.(2018)Daunoraviciene, Adomaviciene, Grigonyte, Gri{\v{s}}kevi{\v{c}}ius, and Juocevicius]{daunoraviciene2018effects}
Kristina Daunoraviciene, Ausra Adomaviciene, Agne Grigonyte, Julius Gri{\v{s}}kevi{\v{c}}ius, and Alvydas Juocevicius.
\newblock Effects of robot-assisted training on upper limb functional recovery during the rehabilitation of poststroke patients.
\newblock \emph{Technology and Health Care}, 26\penalty0 (S2):\penalty0 533--542, 2018.

\bibitem[Desanghere and Marotta(2015)]{desanghere2015influence}
Loni Desanghere and Jonathan~J Marotta.
\newblock The influence of object shape and center of mass on grasp and gaze.
\newblock \emph{Frontiers in psychology}, 6:\penalty0 1537, 2015.

\bibitem[Dixit and Tedla(2019)]{dixit2019effectiveness}
Snehil Dixit and Jaya~Shanker Tedla.
\newblock Effectiveness of robotics in improving upper extremity functions among people with neurological dysfunction: a systematic review.
\newblock \emph{International Journal of Neuroscience}, 129\penalty0 (4):\penalty0 369--383, 2019.

\bibitem[Dobkin and Dorsch(2011)]{dobkin2011promise}
Bruce~H Dobkin and Andrew Dorsch.
\newblock The promise of mhealth: daily activity monitoring and outcome assessments by wearable sensors.
\newblock \emph{Neurorehabilitation and neural repair}, 25\penalty0 (9):\penalty0 788--798, 2011.

\bibitem[Durandau et~al.(2019)Durandau, Farina, As{\'\i}n-Prieto, Dimbwadyo-Terrer, Lerma-Lara, Pons, Moreno, and Sartori]{durandau2019voluntary}
Guillaume Durandau, Dario Farina, Guillermo As{\'\i}n-Prieto, Iris Dimbwadyo-Terrer, Sergio Lerma-Lara, Jose~L Pons, Juan~C Moreno, and Massimo Sartori.
\newblock Voluntary control of wearable robotic exoskeletons by patients with paresis via neuromechanical modeling.
\newblock \emph{Journal of neuroengineering and rehabilitation}, 16:\penalty0 1--18, 2019.

\bibitem[Garc{\'\i}a-Vellisca et~al.(2021)Garc{\'\i}a-Vellisca, Matran-Fernandez, Poli, and Citi]{garcia2021hand}
MA~Garc{\'\i}a-Vellisca, Ana Matran-Fernandez, Riccardo Poli, and Luca Citi.
\newblock Hand-movement prediction from emg with lstm-recurrent neural networks.
\newblock In \emph{2021 Global Medical Engineering Physics Exchanges/Pan American Health Care Exchanges (GMEPE/PAHCE)}, pages 1--5. IEEE, 2021.

\bibitem[Gavas et~al.(2018)Gavas, Roy, Chatterjee, Tripathy, Chakravarty, and Sinha]{gavas2018enhancing}
Rahul~Dasharath Gavas, Sangheeta Roy, Debatri Chatterjee, Soumya~Ranjan Tripathy, Kingshuk Chakravarty, and Aniruddha Sinha.
\newblock Enhancing the usability of low-cost eye trackers for rehabilitation applications.
\newblock \emph{PloS one}, 13\penalty0 (6):\penalty0 e0196348, 2018.

\bibitem[Karamians et~al.(2020)Karamians, Proffitt, Kline, and Gauthier]{karamians2020effectiveness}
Reneh Karamians, Rachel Proffitt, David Kline, and Lynne~V Gauthier.
\newblock Effectiveness of virtual reality-and gaming-based interventions for upper extremity rehabilitation poststroke: a meta-analysis.
\newblock \emph{Archives of physical medicine and rehabilitation}, 101\penalty0 (5):\penalty0 885--896, 2020.

\bibitem[Kim et~al.(2023)Kim, Ji, Hwang, Ryu, Jin, Kim, and Kim]{kim2023three}
Sung-Hoon Kim, Dong-Min Ji, In-Su Hwang, Jinwhan Ryu, Sol Jin, Soo-A Kim, and Min-Su Kim.
\newblock Three-dimensional magnetic rehabilitation, robot-enhanced hand-motor recovery after subacute stroke: A randomized controlled trial.
\newblock \emph{Brain Sciences}, 13\penalty0 (12):\penalty0 1685, 2023.

\bibitem[Kothe et~al.(2024)Kothe, Brunner, Podmore, Mullen, Blankertz, et~al.]{labstreaminglayer}
Christian Kothe, Clemens Brunner, Joshua Podmore, Tim Mullen, Benjamin Blankertz, et~al.
\newblock {LabStreamingLayer}.
\newblock \url{https://github.com/sccn/labstreaminglayer}, 2024.
\newblock Accessed: March 13, 2024.

\bibitem[Kwakkel et~al.(2019)Kwakkel, van Wegen, Burridge, Winstein, Van~Dokkum, Alt~Murphy, Levin, and Krakauer]{kwakkel2019standardized}
G~Kwakkel, Erwin~EH van Wegen, Jane~H Burridge, CJ~Winstein, Liesjet Elisabeth~Henriette Van~Dokkum, Margit Alt~Murphy, MF~Levin, and JW~Krakauer.
\newblock Standardized measurement of quality of upper limb movement after stroke: consensus-based core recommendations from the second stroke recovery and rehabilitation roundtable.
\newblock \emph{Neurorehabilitation and neural repair}, 33\penalty0 (11):\penalty0 951--958, 2019.

\bibitem[Little et~al.(2021)Little, K~Pappachan, Yang, Noronha, Campolo, and Accoto]{little2021elbow}
Kieran Little, Bobby K~Pappachan, Sibo Yang, Bernardo Noronha, Domenico Campolo, and Dino Accoto.
\newblock Elbow motion trajectory prediction using a multi-modal wearable system: A comparative analysis of machine learning techniques.
\newblock \emph{Sensors}, 21\penalty0 (2):\penalty0 498, 2021.

\bibitem[Luo et~al.(2010)Luo, Lim, Yang, Tee, Li, Gu, Nguen, Chen, and Yeo]{luo2010interactive}
Zhiqiang Luo, Chee~Kian Lim, Weiting Yang, Ke~Yen Tee, Kang Li, Chao Gu, Kim~Doang Nguen, I-Ming Chen, and Song~Huat Yeo.
\newblock An interactive therapy system for arm and hand rehabilitation.
\newblock In \emph{2010 IEEE Conference on Robotics, Automation and Mechatronics}, pages 9--14. IEEE, 2010.

\bibitem[Mochizuki et~al.(2019)Mochizuki, Centen, Resnick, Lowrey, Dukelow, and Scott]{mochizuki2019movement}
George Mochizuki, Andrew Centen, Myles Resnick, Catherine Lowrey, Sean~P Dukelow, and Stephen~H Scott.
\newblock Movement kinematics and proprioception in post-stroke spasticity: assessment using the kinarm robotic exoskeleton.
\newblock \emph{Journal of neuroengineering and rehabilitation}, 16:\penalty0 1--13, 2019.

\bibitem[Mubin et~al.(2019)Mubin, Alnajjar, Jishtu, Alsinglawi, and Al~Mahmud]{mubin2019exoskeletons}
Omar Mubin, Fady Alnajjar, Nalini Jishtu, Belal Alsinglawi, and Abdullah Al~Mahmud.
\newblock Exoskeletons with virtual reality, augmented reality, and gamification for stroke patients’ rehabilitation: systematic review.
\newblock \emph{JMIR rehabilitation and assistive technologies}, 6\penalty0 (2):\penalty0 e12010, 2019.

\bibitem[Novak and Riener(2013)]{novak2013enhancing}
Domen Novak and Robert Riener.
\newblock Enhancing patient freedom in rehabilitation robotics using gaze-based intention detection.
\newblock In \emph{2013 IEEE 13th international conference on rehabilitation robotics (ICORR)}, pages 1--6. IEEE, 2013.

\bibitem[Park et~al.(2019)Park, Jeon, Lim, Koo, Lee, Kim, Lee, Song, and Hong]{park2019feasibility}
Jeong~Hye Park, Han~Jae Jeon, Eun-Cheon Lim, Ja-Won Koo, Hyo-Jeong Lee, Hyung-Jong Kim, Jung~Seop Lee, Chang-Geun Song, and Sung~Kwang Hong.
\newblock Feasibility of eye tracking assisted vestibular rehabilitation strategy using immersive virtual reality.
\newblock \emph{Clinical and experimental otorhinolaryngology}, 12\penalty0 (4):\penalty0 376, 2019.

\bibitem[Saggio et~al.(2015)Saggio, Riillo, Sbernini, and Quitadamo]{saggio2015resistive}
Giovanni Saggio, Francesco Riillo, Laura Sbernini, and Lucia~Rita Quitadamo.
\newblock Resistive flex sensors: a survey.
\newblock \emph{Smart Materials and Structures}, 25\penalty0 (1):\penalty0 013001, 2015.

\bibitem[Sharma et~al.(2017)Sharma, Sharma, and Athaiya]{sharma2017activation}
Sagar Sharma, Simone Sharma, and Anidhya Athaiya.
\newblock Activation functions in neural networks.
\newblock \emph{Towards Data Sci}, 6\penalty0 (12):\penalty0 310--316, 2017.

\bibitem[Tarnita et~al.(2022)Tarnita, Geonea, Pisla, Carbone, Gherman, Tohanean, Tucan, Abrudan, and Tarnita]{tarnita2022analysis}
Daniela Tarnita, Ionut~Daniel Geonea, Doina Pisla, Giuseppe Carbone, Bogdan Gherman, Nicoleta Tohanean, Paul Tucan, Cristian Abrudan, and Danut~Nicolae Tarnita.
\newblock Analysis of dynamic behavior of parreex robot used in upper limb rehabilitation.
\newblock \emph{Applied Sciences}, 12\penalty0 (15):\penalty0 7907, 2022.

\bibitem[Thostenson et~al.(2019)Thostenson, Doshi, and Chaudhari]{thostenson2019flexible}
Erik~T Thostenson, Sagar Doshi, and Amit Chaudhari.
\newblock Flexible and wearable sensors for human motion analysis.
\newblock In \emph{Abstracts of Papers of the American Chemical Society}, volume 258. AMER CHEMICAL SOC 1155 16TH ST, NW, Washington, DC 20036 USA, 2019.

\bibitem[Trigili et~al.(2019)Trigili, Grazi, Crea, Accogli, Carpaneto, Micera, Vitiello, and Panarese]{trigili2019detection}
Emilio Trigili, Lorenzo Grazi, Simona Crea, Alessandro Accogli, Jacopo Carpaneto, Silvestro Micera, Nicola Vitiello, and Alessandro Panarese.
\newblock Detection of movement onset using emg signals for upper-limb exoskeletons in reaching tasks.
\newblock \emph{Journal of neuroengineering and rehabilitation}, 16:\penalty0 1--16, 2019.

\bibitem[Vibhuti et~al.(2023)Vibhuti, Kumar, and Kataria]{vibhuti2023efficacy}
Vibhuti, Neelesh Kumar, and Chitra Kataria.
\newblock Efficacy assessment of virtual reality therapy for neuromotor rehabilitation in home environment: a systematic review.
\newblock \emph{Disability and Rehabilitation: Assistive Technology}, 18\penalty0 (7):\penalty0 1200--1220, 2023.

\bibitem[Walker et~al.(2023)Walker, Phung, Chakraborti, Williams, and Szafir]{walker2023virtual}
Michael Walker, Thao Phung, Tathagata Chakraborti, Tom Williams, and Daniel Szafir.
\newblock Virtual, augmented, and mixed reality for human-robot interaction: A survey and virtual design element taxonomy.
\newblock \emph{ACM Transactions on Human-Robot Interaction}, 12\penalty0 (4):\penalty0 1--39, 2023.

\bibitem[Wei and Wu(2023)]{wei2023application}
Suyao Wei and Zhihui Wu.
\newblock The application of wearable sensors and machine learning algorithms in rehabilitation training: A systematic review.
\newblock \emph{Sensors}, 23\penalty0 (18):\penalty0 7667, 2023.

\bibitem[Wonsick and Padir(2020)]{wonsick2020systematic}
Murphy Wonsick and Taskin Padir.
\newblock A systematic review of virtual reality interfaces for controlling and interacting with robots.
\newblock \emph{Applied Sciences}, 10\penalty0 (24):\penalty0 9051, 2020.

\bibitem[Yang et~al.(2023)Yang, Garg, Gao, Yuan, Noronha, Ang, and Accoto]{yang2023learning}
Sibo Yang, Neha~P Garg, Ruobin Gao, Meng Yuan, Bernardo Noronha, Wei~Tech Ang, and Dino Accoto.
\newblock Learning-based motion-intention prediction for end-point control of upper-limb-assistive robots.
\newblock \emph{Sensors}, 23\penalty0 (6):\penalty0 2998, 2023.

\end{thebibliography}

\end{document}